\documentclass[traditabstract]{aa}
\usepackage{amsmath,amssymb}
\usepackage{txfonts}
\usepackage{graphicx}
\usepackage{color}
\usepackage{url}
\usepackage{natbib}

\newcommand{\eg}{{{e.g.~}}}
\newcommand{\egc}{{{e.g.}}}
\newcommand{\ie}{{i.e.~}}
\newcommand{\iec}{{i.e.,~}}
\begin{document}

\title{The solar magnetic field since 1700}
\subtitle{I. Characteristics of sunspot group emergence and
reconstruction of the butterfly diagram}

\author{J.~{Jiang}\and R.~H.~{Cameron}\and D.~{Schmitt}\and M.~{Sch\"ussler}}
\institute{Max-Planck-Institut f\"ur Sonnensystemforschung, 37191 Katlenburg-Lindau, Germany}
   
\date{Received -- ; accepted --}

\begin{abstract}
{We use the historic record of sunspot groups compiled by the Royal
Greenwich Observatory together with  the sunspot number to
derive the statistical properties of sunspot group emergence in
dependence of cycle phase and strength. In particular we discuss the
latitude, longitude, area and tilt angle of sunspot groups as
functions of the cycle strength and of time during the solar cycle.

Using these empirical characteristics  the time-latitude diagram of sunspot group
emergence (butterfly diagram) is reconstructed from 1700 onward
on the basis of the  Wolf and group sunspot numbers.
This reconstruction will be useful in studies of the long-term
evolution of the Sun's magnetic field.

}
\end{abstract}

\maketitle

\section{Introduction}
The synoptic record of sunspot emergence is an important input into, for example,
long term reconstructions of the solar open flux \citep[\eg][]{Cameron10},
solar irradiance variations \citep[\eg][]{Crouch08} and is 
relevant for understanding the solar dynamo \citep[for a recent review see][]{Charbonneau10}. 
The quantities which are used in these types
of studies include the sunspot areas, emergence latitudes and longitudes
as well as the tilt angle between sunspots of opposite polarities within a group. 

As an example of such a study,  in \cite{Cameron10} we used the observed Royal Greenwich 
Observatory\footnote{All data was obtained from the NOAA website \url{http://www.ngdc.noaa.gov/stp/solar/solardataservices.html}}  
(RGO) records of sunspot areas, longitudes, latitudes and the 
Mount Wilson Observatory and Kodaikanal records of tilt angles  \citep{Howard84, Howard99, Sivaraman93} as the input to a surface flux transport 
model \citep[\eg][]{Devore85,Sheeley85,Wang89,Baumann04}. 
The results from the model was the large-scale evolution of the surface magnetic field, which  was extrapolated
into interplanetary space using a current sheet source surface (CSSS) extrapolation \citep[\eg][]{Zhao95}. The open
field calculated from the model was then compared to that inferred from observations of the geomagnetic
aa index. The time period analyzed was restricted by the fact that the RGO dataset only extends back to
1874 and the MWO and Kodaikanal tilt angle datasets are even shorter.

The purpose of this paper is to construct partially synthetic datasets of sunspot 
emergence covering the period from 1700 to 2010. In the second paper in this series
we intend to use these semi-synthetic records with the surface flux transport model
and CSSS extrapolation. The semi-synthetic data sets however have a much wider application,
for example in irradiance studies and in understanding the solar dynamo. We therefore
here present the analysis and methods for creating them.

The time dependence of the emergence is taken from the sunspot number data,
either the monthly group sunspot number, $R_G$ \citep{Hoyt98}, or the monthly Wolf sunspot number, $R_Z$ \citep{Wolf61}.
Correlations between the strength of the cycle, derived from $R_G$ or $R_Z$, and the
areas, emergence latitudes and longitudes, and tilt angles of sunspot groups are sought. 
For this purpose we use the Royal Greenwich Observatory (RGO) record of sunspot group areas, latitudes, 
longitudes as well as the MWO and Kodaikanal records of sunspot group tilt angles. 
Unlike previous studies \citep[\eg][]{Li03, Solanki08, Dasi-Espuig10, Ivanov10}, we consider  
correlations of many of the emergence properties with cycle properties derived from the monthly sunspot number.

These correlations are then used in conjunction with the $R_G$ and $R_Z$ records 
to construct artificial sunspot group data extending back to 1700. 
As the time dependence is taken from observations and the other properties of the sunspot
groups are synthetic, the constructed timeseries is semi-synthetic.
Since the correlations are only statistical in nature, the individual reconstructions 
are realizations drawn randomly from a population with the observed statistics. 
This enables Monte-Carlo-type studies on the longer term evolution of, \eg the Sun's open flux, 
polar fields and irradiance variations.

The paper is structured as follows: in Section 2 we describe the
datasets and ways to define various cycle parameters such as cycle
strength, starting time and length of each cycle. In Section 3 we
discuss the correlations between the spatial distribution and the
properties of the cycle as determined from $R_G$. 
In Section 4 we use these correlations  to  
reconstruct the butterfly diagram from 1700 onwards.

\section{Cycle parameters determined from sunspot numbers} \label{s:defs}

\subsection{Sunspot number datasets}
Monthly values of the Wolf sunspot number, $R_Z$, are  available from 1749
onward and yearly values reach back to 1700.
Monthly values of $R_Z$ before 1749 can be estimated by interpolation of yearly
$R_Z$ values. The group sunspot number $R_G$ extends further back in time, to 1610,
and is again interpolated to obtain monthly values when the dataset is incomplete.
As described by \cite{Usoskin08} and \cite{Hathaway10} the two datasets have different definitions and
depend on different combinations of solar observations. 

\begin{figure}
\centering
\resizebox{\hsize}{!}{\rotatebox{90}{\includegraphics{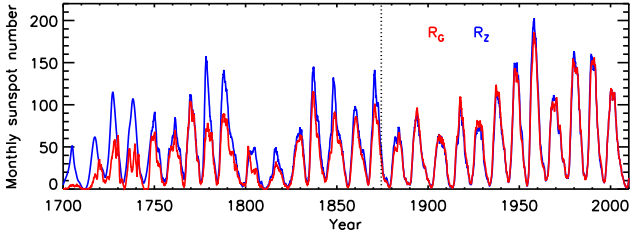}}}
\caption{Time evolution of monthly Wolf sunspot number $R_Z$ (blue
line) and group sunspot number $R_G$ (red line). The dotted vertical line
denotes the year 1874, after which the two sunspot number
datasets are nearly identical. Note that $R_G$ ends in 1995, after which it is assumed to
be equal to $R_Z$.} \label{fig:WolfVSgroup}
\end{figure}

Figure \ref{fig:WolfVSgroup} shows the two sunspot numbers over the period from 1700 to 2010.
After the 1870s they are nearly identical. Since this covers the 
period of more detailed RGO sunspot data, the two data sets are almost equivalent
for use in determining the empirical correlations with the RGO data. We have (arbitrarily) chosen to
use $R_G$. Differences however are to be expected in the reconstructed sunspot group 
data prior to 1874, as will be seen in Section 4.

\subsection{Cycle parameters}
On the basis of the $R_G$ data we define three parameters for each cycle. 
Two of these parameters concern  the strength of a cycle.
The first, $S_n$, is the maximum of the 12 month running mean of $R_G$.
The second, $\tilde{S}_n$, is the sum of $R_G$ over the cycle. 

We also need information as to the timing of the cycle, and for this we use the 
time of the solar minima, $t_\mathrm{min}$ \citep{Harvey99}, which we take from the NGDC
website.
These times and cycle strengths are listed in Table~\ref{tbl:cys}. 

\begin{table}
\caption{Parameters for solar cycles $-$4 to 23 (from 1700 onward) derived from the group
sunspot number $R_G$.} \label{tbl:cys}
\centering
\begin{tabular}{cccc}
\hline
cycle no. & $t_\mathrm{min}$ & $\tilde{S}_n/100$ & $S_n$ \\
\hline
$-$4 & 1698.0 &   3 &   5  \\
$-$3 & 1712.0 &  16 &  35  \\
$-$2 & 1723.5 &  33 &  64  \\
$-$1 & 1734.0 &  31 &  53  \\
 0 & 1745.0 &  36 &  67  \\
 1 & 1755.2 &  43 &  70  \\
 2 & 1766.5 &  61 & 103  \\
 3 & 1775.5 &  45 &  78  \\
 4 & 1784.7 &  70 &  89  \\
 5 & 1798.3 &  23 &  51  \\
 6 & 1810.6 &  19 &  32  \\
 7 & 1823.3 &  41 &  64  \\
 8 & 1833.9 &  60 & 116  \\
 9 & 1843.5 &  64 &  91  \\
10 & 1856.0 &  55 &  85  \\
11 & 1867.2 &  57 & 101  \\
12 & 1878.9 &  40 &  67  \\
13 & 1889.6 &  57 &  96  \\
14 & 1901.7 &  46 &  64  \\
15 & 1913.6 &  59 & 109  \\
16 & 1923.6 &  56 &  81  \\
17 & 1933.8 &  77 & 123  \\
18 & 1944.2 &  86 & 143  \\
19 & 1954.3 & 105 & 186  \\
20 & 1964.9 &  83 & 108  \\
21 & 1976.5 & 100 & 154  \\
22 & 1986.8 &  92 & 156  \\
23 & 1996.4 &  81 & 119  \\
\hline
\end{tabular}
\end{table}

\section{Characteristics of sunspot group emergence derived from the RGO sunspot data}\label{s:char}

In this section we discuss the empirical relationships between 
the strength of the cycle given in Table 1 and the latitudes, 
longitudes and areas of sunspot groups as recorded in the RGO dataset.
The RGO records cover the period from 1874 to 1976 (cycles 12 to 20).
The sunspot groups are considered at the times of their 
maximum reported area. 
 
Correlations between the strength of the cycle and the tilt angle of sunspot groups 
rely on the MWO and Kodaikanal data as discussed in \citet{Dasi-Espuig10}.
Some properties, such as the average latitude at which sunspots emerge, 
vary throughout the solar cycle. In this case it is important to consider
the data from different cycles at the same phase. We then look for correlations at
a fixed phase through the cycle, where the cycle is taken to begin and end at adjacent activity
minima.

\subsection{Latitude distribution}\label{s:mean_lati}

The latitude distribution of sunspots is the clearest example of where it is necessary to 
consider the phase during the cycle: early in the cycle sunspots appear
at higher latitudes than later in the cycle. 

\begin{figure}
\centering
\resizebox{\hsize}{!}{\includegraphics[width=0.5\textwidth]{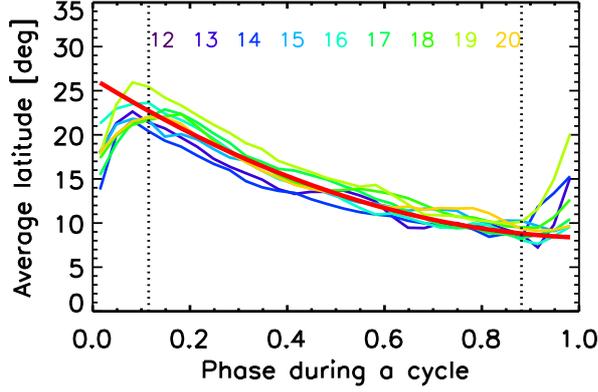}}
\caption{Average latitude of sunspot groups at different cycle phases. Colors indicate
different cycles (cycles 12--20). The thick red curve shows the polynomial fit
given by Equation~(\ref{eq:lati_tmin}) to all cycles. The two vertical dotted lines indicate
the times when the overlapping of cycles strongly affects the average latitudes.
} \label{fig:lati_tmin}
\end{figure}

We break the time between adjacent minima into 30 equal phases.  
For each cycle, $n$, we can then calculate the  mean latitude, ${{\lambda}_n^{i}}$,
averaged over the $i^{\mathrm{th}}$ phase bin.
Figure \ref{fig:lati_tmin} shows
${{\lambda}_n^{i}}$ for cycles 12 to 20 as a function of
the phase of the cycle.
Because the cycles partially overlap, the
first 3 phase bins near the start and the last 3 near the end of a
cycle show a mixture of spots from adjacent cycles. This mixture is of
lower-latitude spots from the end of a cycle and
higher-latitude spots from the beginning of the subsequent cycle. The
average latitude during these initial and final phases is thus
difficult to interpret. 

The phase average ${\lambda}^{i}$ over all cycles defined as
$\displaystyle\frac{1}{9}\sum_{n=12}^{20}{\lambda}_n^{i}$
is well fit by a second degree polynomial:
\begin{equation}
\label{eq:lati_tmin}
{\lambda}^i=26.4-34.2(i/30)+16.1(i/30)^{2} 
\end{equation}
over the range  $4 \le i\le 27$. The rms difference between the fit and the mean latitudes is 0.3$^{\circ}$.

\begin{figure}
\centering
\resizebox{\hsize}{!}{\includegraphics[width=1.\textwidth]{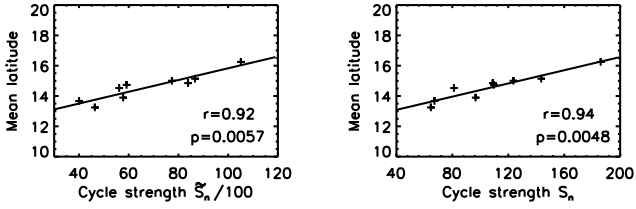}}
\caption{Correlation between cycle averaged latitudes ${\lambda}_n$
and cycle strength defined by the total sunspot number ($\tilde{S}_n$, left
panel) and the maximum sunspot number ($S_n$, right panel), respectively.}
\label{fig:aveLat_str}
\end{figure}

As previously reported \citep{Li03, Solanki08}, there is a strong correlation between the 
strength of the cycle and the average latitudes of emergence. To evaluate this correlation
we calculated the mean latitude of emergence as 
\begin{equation}
{\lambda}_n=\displaystyle\frac{1}{24}\sum_{i=4}^{27}{\lambda}_n^{i}.
\label{eq:lati_ave}
\end{equation}
Figure \ref{fig:aveLat_str} shows the relation between the average
latitudes ${\lambda}_n$ and cycle strengths defined by
$\tilde{S}_n$ (left panel) and $S_n$ (right panel). 
In both cases, a statistically significant ($p$-value $<0.05$) correlation coefficient higher 
than 0.9 is found. The correlations between $S_n$ and ${\lambda}_n$ (r=0.94)  
and between $\tilde{S}_n$ and ${\lambda}_n$ (r=0.92) are similar. We hereafter 
focus on $S_n$. The linear fit for the ${\lambda}_n$ is given by
\begin{equation}
\label{eq:lati_str} {\lambda}_n=12.2+0.022S_n.
\end{equation}

\begin{figure}
\centering
\resizebox{\hsize}{!}{\includegraphics[width=1.0\textwidth]{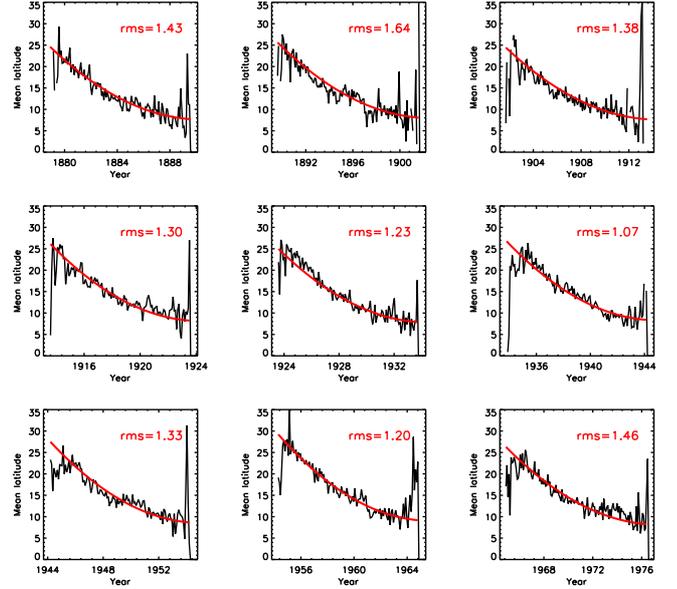}}
\caption{Comparison of the mean latitudes of sunspot emergence for cycles 12--20
between the observation (black line, monthly average) and the
fit (red line). 
The rms deviations (in degrees) between observation
and reconstruction are given excluding the first and the last two
years of a cycle.} \label{fig:lati_analy_t0tmin}
\end{figure}

This observed correlation can be used with Equation~(\ref{eq:lati_ave})
to model the phase dependence of the mean latitude of emergence for different cycles:
\begin{equation}
\label{eq:lati_tmin_2}
{\lambda}^i_n=(26.4-34.2(i/30)+16.1(i/30)^{2})({\lambda}_n/{\langle{\lambda_n}\rangle_{12-20}}) 
\end{equation}
for $1 \le i \le 30$ and 
where $\langle{\lambda}\rangle_{12-20}=14.6^{\circ}$ is the average latitude of sunspot emergence over all the cycles.
The fit to each cycle is shown in Figure~\ref{fig:lati_analy_t0tmin}.
The mean rms deviation between observation and reconstruction, excluding the first and last two years of each cycle,  
is 1.33$^\circ$. 

\subsection{Width of the latitude distribution}
\begin{figure}
\centering
\resizebox{\hsize}{!}{\includegraphics[width=0.45\textwidth]{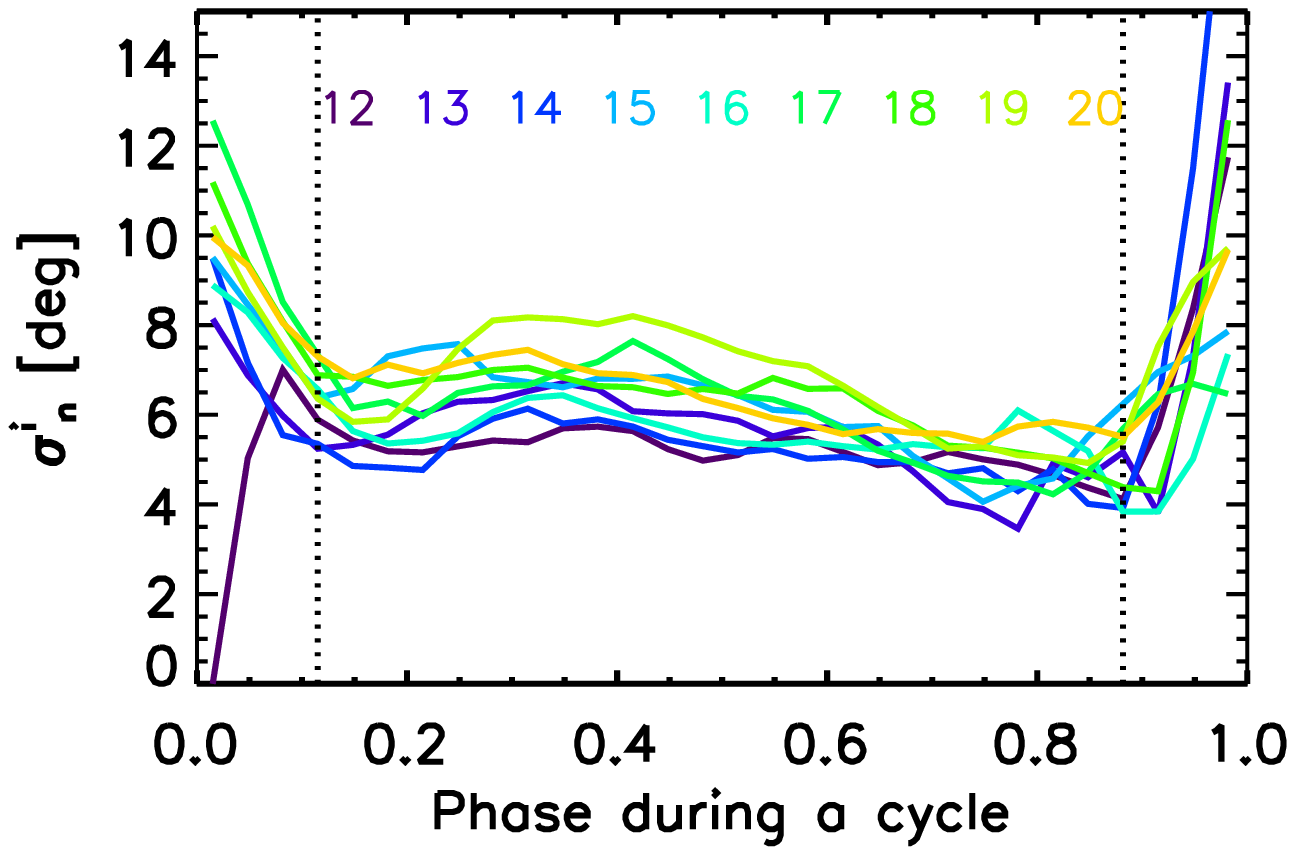}}
\resizebox{\hsize}{!}{\includegraphics[width=0.45\textwidth]{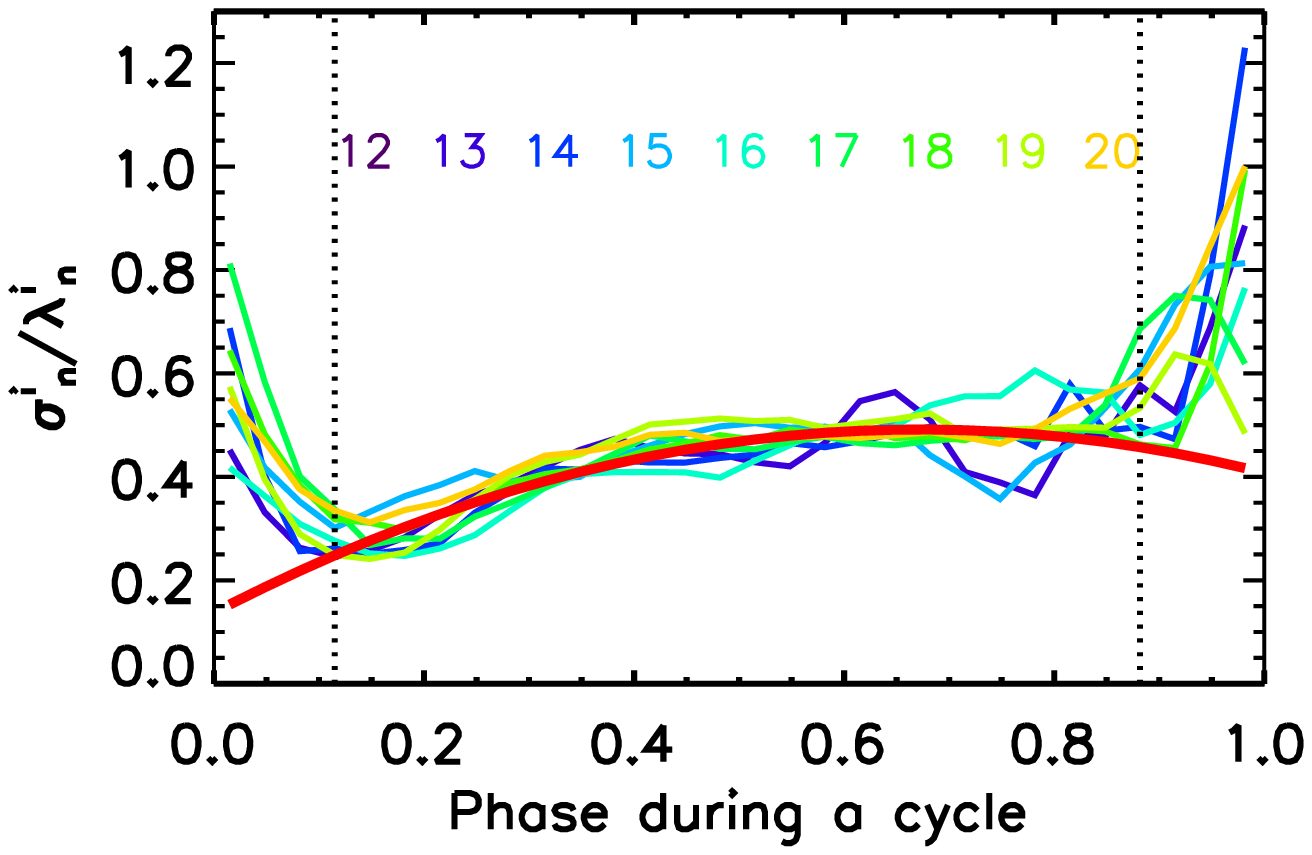}}
\caption{Upper panel: standard deviation $\sigma_n^i$ of the
latitudinal distribution of phase-binned latitudes for cycles
12--20. The line style is the same as in Figure \ref{fig:lati_tmin}.
Lower panel: similar except $\sigma_n^i/\lambda_n^i$ is shown.
The thick red curve shows the polynomial fit given by Equation~(\ref{eq:lati_tmin_w}). }
\label{fig:lati_tmin_sca}
\end{figure}

Sunspots emerge over a range of latitudes at any phase of the cycle.
We consider the standard 
deviation, $\sigma_n^{i}$, of the latitudinal distribution during
phase $i$ of cycle $n$. The upper panel in Figure \ref{fig:lati_tmin_sca} shows 
$\sigma_n^{i}$ for cycles $n=12-20$. A tighter relationship is found if 
we consider the ratio $\sigma_n^{i}/\lambda_n^{i}$ which is 
shown in the lower panel of the same figure. 

A second-order polynomial fit 
\begin{equation}
\label{eq:lati_tmin_w} \sigma^i=(0.14+1.05(i/30)-0.78(i/30)^{2})\lambda^i
\end{equation}
matches the data well. In our semi-synthetic reconstructions we assume
a Gaussian distribution with a half width of $\sigma_n^i$ and exclude
points deviating from the mean by  more then $2\sigma_n^i$.

\subsection{Longitude distribution}

The emergence longitudes of sunspot groups is known to be not entirely random
\citep[\eg][]{Bumba65,Bai88,Berdyugina03, Zhang07}.
\citet{Castenmiller86} introduced the term `sunspot nests' to
describe the tendency for sunspots to appear near where other spots.  
Up to 30\% of the sunspot groups have previously been found to be associated with such nests. 

Our motivation for considering the longitudes distribution is, for example, that the open flux of the 
Sun during activity maxima is dominated by the equatorial low order multipoles 
\citep[see, \eg][]{Cameron10}. The strengths of these multipoles, and the dipole 
in particular, depend on how randomly or systematically the sunspots appear in longitude.
We therefore begin by considering the equatorial dipole moment of an 
individual sunspot group. The first step is to convert the sunspot area, $A$, in the
RGO dataset into magnetic flux. Here we follow 
\cite{van_Ballegooijen98} and \cite{Baumann04} and take the total flux of
the group to be proportional to the area of the active region (sunspot area and
plage area)
\begin{equation}
A_R=A+414 +21A-0.0036A^2
\end{equation}
where all values are in $\mu$Hem \citep{Chapman97}. 
The equatorial dipole moment of the sunspot group is then assumed to be 
proportional to the area of the group, $A_R$, multiplied
by the separation between the opposing polarities which we 
take to be proportional to $A_R^{1/2}$. 
For a sunspot group $j$ in the northern hemisphere and near the 
equator, with area $A_{R,j}$ and central
longitude $\phi_j$, the axis of the equatorial dipole is orientated in the direction
$\phi_j \pm 90^{\circ}$. The component of the dipole in the equatorial
plane in direction $\phi$ is thus proportional to $A_{R,j}^{3/2} \cos(\phi-\phi_j)$. 
For a number of sunspots groups in the northern hemisphere, the component
of the resulting dipole moment in the direction $\phi$ is proportional to 
\begin{equation}
\sum_{\mathrm{north}} A_{R,j}^{3/2} \cos(\phi-\phi_j).
\end{equation}
Since the sunspot groups in the southern hemisphere have the opposite polarity 
orientation, the corresponding component of the dipole moment is proportional to
\begin{equation}
-\sum_{\mathrm{south}} A_{R,j}^{3/2} \cos(\phi-\phi_j).
\end{equation} 
Therefore dipole moment of all the spot groups from both hemispheres is proportional to 
\begin{equation}
m=\sum_{\mathrm {north}} A_{R,j}^{3/2} \cos(\phi-\phi_j)-\sum_{\mathrm{south}} A_{R,j}^{3/2} \cos(\phi-\phi_j).
\end{equation}
Clearly, $m$ depends on the sunspots which are included in the sum as well as the direction $\phi$. We define 
$m(t,\tau,\phi)$ to include in the sum all spots emerging between times $t$ and $t+\tau$.

We are not able to reproduce the longitudes at
which sunspots have appeared from 1700 onwards or even those longitudes
where nesting has occurred. Our aim is to obtain statistical information
about the degree of nesting by measuring the degree of
non-randomness present in the RGO data. We do this by 
creating three copies of the RGO records. In the first copy
we replace the observed longitudes with randomly chosen longitudes from
a uniform distribution. We use the subscript notation $m_{\mathrm{ran}}$
for this dataset. The second copy 
has the longitudes of its spots changed so that they all appear at 0$^{\circ}$
in the northern hemisphere and 180$^{\circ}$ in the southern hemisphere. This
choice maximizes the equatorial magnetic dipole moment since sunspot groups 
in opposite hemispheres have opposite polarity orientations in accordance with Hale's
law.  We use the subscript notation $m_{\mathrm{ord}}$ for this dataset. 
The third copy retains the observed longitudes and uses the subscript 
notation $m_{\mathrm{obs}}$.

\begin{figure}
\centering
\resizebox{\hsize}{!}{\includegraphics[width=0.6\textwidth]{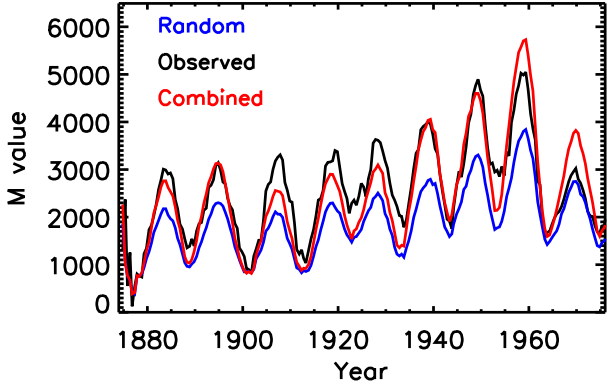}}
\caption{Time evolution of the proxy for the equatorial dipole moment, $M$, 
during 1874--1976 for $\tau=6$~months. The random (blue curve), observed (black curve) and 
combined (red curve) models for the longitude distribution of sunspot emergence are shown.
The value of $c=0.15$ corresponds to  $\tau=6$~months.  }
\label{fig:M_vs_time}
\end{figure}

\begin{figure}
\centering
\resizebox{\hsize}{!}{\includegraphics[width=0.6\textwidth]{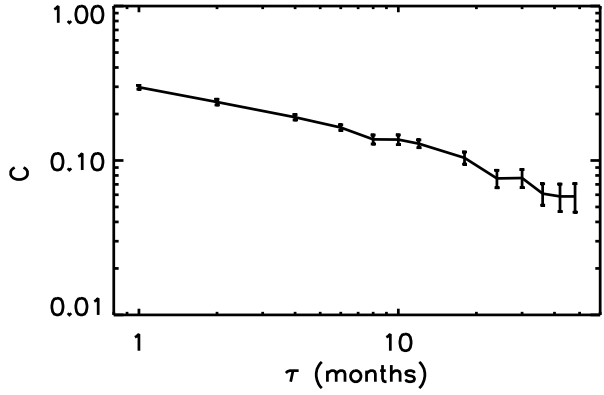}}
\caption{Degree of non-randomness, $\mathrm{c}$, as a function of the 
period over which we take the sunspots, $\tau$. Error bars represent the
standard deviation resulting from 20 sets of random longitudes to
obtain $M_{\mathrm{ran}}$.} \label{fig:C_vs_DeltaT}
\end{figure}

For a given $\tau$ we measure the amount of nonrandomness, $\mathrm{c}(\tau)$. We
assume that the magnitude of the observed sunspot dipole moment 
\begin{equation}
M_{\mathrm{obs}}(t,\tau) \equiv \max_{\phi}\{m_{\mathrm{obs}}(t,\tau,\phi)\}
\end{equation}
is the magnitude of a linear combination of the random and ordered datasets
\begin{equation}
M_{\mathrm{c}}(t,\tau)\equiv \max_{\phi}\{\mathrm{c}(\tau) m_{\mathrm{ord}}(t,\tau)+[1-\mathrm{c}(\tau)] m_{\mathrm{ran}}(t,\tau)\}
\end{equation}
and seek the $\mathrm{c}(\tau)$ which minimizes the rms differences. 
Figure~\ref{fig:M_vs_time} shows $M_{\mathrm{obs}}$, $M_{\mathrm{ran}}$ and $M_{\mathrm{c}}$
for the example $\tau=6$~months. 
The difference between $M_{\mathrm{obs}}$ and $M_{\mathrm{ran}}$ during activity maxima reflects
the amount of nesting, \iec the nonrandom longitude distribution.
The similarity of all the curves near the minima is a consequence of the fact that,
when there are few spot groups, they are automatically highly ordered.
We show the dependence of $\mathrm{c}$ on $\tau$ in Figure~\ref{fig:C_vs_DeltaT}.

For different studies the appropriate value of $\tau$ will vary. For irradiance studies $\tau=$1 month
would seem to be an appropriate choice because it is the instantaneous clustering of the sunspots
which is important. For surface flux transport simulations, $\tau=6$~months is more relevant as this is 
approximatley the time it takes for the emerging flux to be sheared by differential rotation
\citep[][p.~162]{Schrijver00}. 

\subsection{Area distribution}
\label{s:area_dis}
In this section we consider the area distribution of sunspot groups
based upon the RGO dataset. As previously stated, we consider each group only at the 
time when it has its maximum recorded area.  We begin by
considering the area distribution of the entire dataset. We will then
look at the partially related questions of its dependence on the phase 
during the cycle and on the latitudes at which the spots appear.

\subsubsection{Area distribution function}

\begin{figure}
\centering
\resizebox{\hsize}{!}{\includegraphics[width=0.5\textwidth,angle=90]{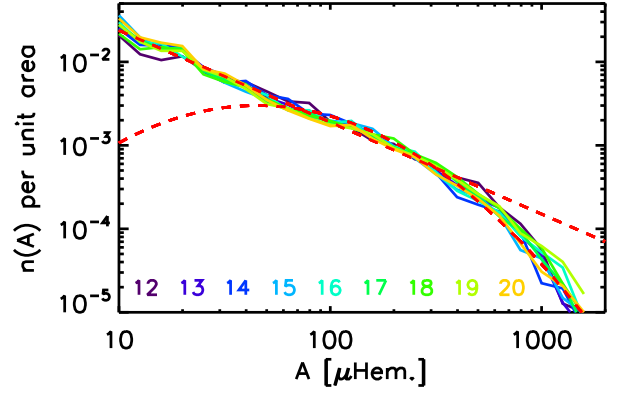}}
\caption{Number density function of the group areas for each cycle of RGO area dataset.
The two red dashed curves show the fits from Equations~(\ref{eq:n_A1}) and (\ref{eq:n_A2}). } 
\label{fig:size_func_cycs}
\end{figure}

Figure~\ref{fig:size_func_cycs} shows the number
density function of sunspot group areas. The behaviour is approximately
a power law below $300\, \mu$Hem with a turnover to an almost 
log-normal distribution above \citep[see also][]{Zhang10}. 
There is a large range (from $60$ to  $300\, \mu$Hem)
where both functional forms are good approximations to the data.
The fits for the two sections of the curves are 
\begin{equation}
\label{eq:n_A1} n(A)=0.3 A^{-1.1} \,\, \mathrm{ for }\, A<300\,\mu{\mathrm{Hem}}  
\end{equation}
and 
\begin{equation}
\label{eq:n_A2} n(A)=0.003\exp[-\frac{1}{2\ln 3}(\ln A-\ln 45)^2] \,\, \mathrm{ for }\, A>60\,\mu{\mathrm{Hem}}.
\end{equation}
The differences between the two fits are mainly in the tails, which is
a partial explanation of why both log-normal and power-law distributions
have been reported in the past 
\citep[see, \eg][]{Bogdan88, Harvey93, Baumann05, Harvey93, Schrijver94}.
We also comment that we are here considering the sunspot group areas from the RGO
record, which includes both the umbral and penumbral area but excludes the
area of the plage. This also partially explains why our results can differ
from those of previous authors.

\subsubsection{Cycle phase dependence of area distribution}\label{s:area}

\begin{figure}
\centering
\resizebox{\hsize}{!}{\includegraphics[width=0.6\textwidth]{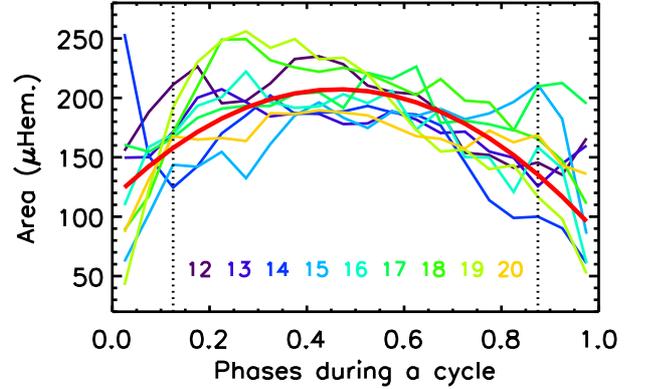}}
\caption{Mean sunspot group area for cycles 12--20. The red curve is 
the fit given by Equation~(\ref{eq:area_time_tmin}).} \label{fig:area_time_tmin}
\end{figure}

We next consider dependence on the cycle phase and latitude of the
area distribution.
For the phase-of-cycle dependence we use the same type of analysis as in 
Section \ref{s:mean_lati}. Figure
\ref{fig:area_time_tmin} shows the area distribution for different cycles as a
function of the phase $1 \le i \le 30$. The average value over all cycles can be fitted by the
second degree polynomial
\begin{equation}
\label{eq:area_time_tmin} A_i=115+396 (i/30)-426 (i/30)^{2}
\end{equation}

\begin{figure}
\centering
\resizebox{\hsize}{!}{\includegraphics[width=0.6\textwidth]{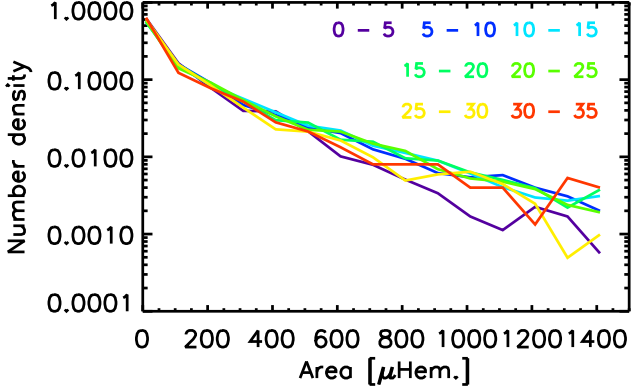}}
\caption{Number density function for 5 degree-binned latitudes.}\label{fig:lati_por}
\end{figure}

There is also a (possibly related) dependence of the areas on latitude.
Figure \ref{fig:lati_por} shows the number density
function of sunspot group areas for 5 degree binned latitudes.
Even after averaging the data in this way there is still 
some scatter apparent in the data. The relative scatter can, which can
be judged from the latitude-to-latitude variation in the plot,  
increases with area for each latitude bin. 
This is because there are relatively 
few large sunspot groups. 
The figure also indicates that large sunspots rarely occur at low ($<5$) latitudes.
This partly reflects the phase dependence of the area distribution.

\subsection{Tilt angle distribution}

Joy's law \citep{Hale19} states that the line joining the centres of the positive and negative
polarities of sunspot groups is systematically tilted with  respect to
the East-West direction. Recently, \citet{Dasi-Espuig10} found that the tilt angles of sunspot groups,  
obtained from the Mount Wilson Observatory and from Kodaikanal observations, show a
cycle-to-cycle variation. They further showed that the average
tilt angle is negatively correlated with the strength of the cycle, \ie the
tilt angles tend to be smaller for stronger cycles. Incorporating the cycle-dependent
tilt angles of sunspot groups in a surface flux transport model,
\citet{Cameron10} reproduced the empirically derived time evolution of the
solar open magnetic flux and the reversal times of the polar fields from 1913
to 1986 (cycles 15--21), here we consider only the time period covered by the 
RGO data and hence omit cycle 21 from our analysis.
The precise dependence of the tilt angle on latitude
is uncertain and it seems that a square root profile matches the data somewhat
better than the usually assumed linear relationship. In this paper, as in CJSS10,
 we consider the square root profile
$\alpha_n=T_n\sqrt{|\lambda|}$, where $\alpha_n$ is the average tilt angle and 
$T_n$ is the constant of proportionality for cycle $n$.

\begin{figure}
\centering
\resizebox{\hsize}{!}{\includegraphics[width=1.0\textwidth]{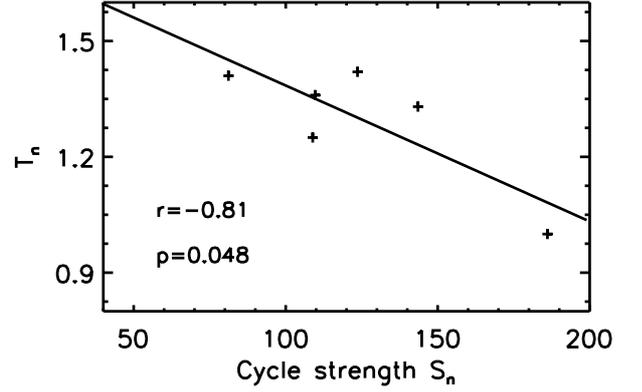}} \caption{The
relationship between cycle strength $S_n$ and $T_n$. } \label{fig:tl}
\end{figure}

For each cycle we determined $T_n$ in a similar way as in
\citet{Dasi-Espuig10} and \citet{Cameron10}. Figure \ref{fig:tl}
shows the correlation between cycle strength $S_n$ and $T_n$. 
The correlation coefficient for $S_n$ is 0.81, slightly higher than
that of  $\tilde{S}_n$ which is 0.78. Under either definition, the cycle strength 
and $T_n$ are significantly correlated ($p<0.05$). 
The linear fit between $S_n$ and $T_n$ is
\begin{equation}
\label{eq:tl} T_n=1.73-0.0035 S_n
\end{equation}

\subsection{Number of sunspot groups as a function of time}
\begin{figure}
\centering
\resizebox{\hsize}{!}{\includegraphics[width=0.6\textwidth,angle=0]{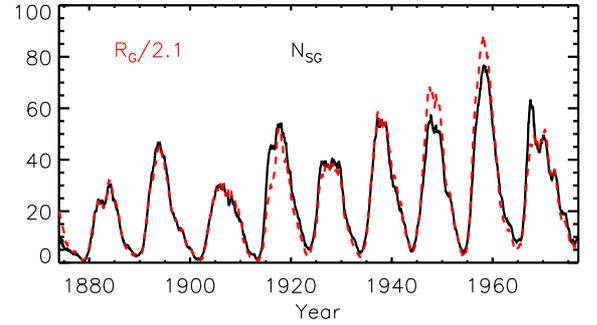}}
\caption{ A comparison of the number of sunspot groups appearing each month in the RGO
data, $N_{SG}$, (black curve) and fit based upon $R_G$ (red dashed curve). 
}\label{fig:sn_sg}
 \end{figure}

Our aim in this paper is to construct semi-synthetic sunspot group records, based on $R_G$
and $R_Z$, 
with similar statistics to those of the RGO data.  We have used $R_G$ to determine strength of 
each cycle and have found correlations which allow us to construct synthetic latitudes, longitudes, 
areas and tilt angles for each spot group. We here determine how many sunspot groups
should appear each month to make the semi-synthetic record have similar statstics as the RGO 
dataset. 
For the period covered by the RGO records, the monthly number should be approximately
the same as the number of groups in the RGO dataset, $N_{SG}$. We have found that
the fit $R_G/2.1$, shown in Figure~\ref{fig:sn_sg}, matches the data well. We use
this fit to reconstruct the number of sunspot groups appearing each month from 1700 onwards.

\section{Reconstruction of sunspot group emergence}

Our goal is to reconstruct sunspot group emergence back to 1700
based solely on sunspot numbers. For that we use the derived
relationships between sunspot groups (mean latitude, latitude width,
longitude distribution, area distribution and tilt angles) and
activity indices (sunspot numbers, strength, and phase of
the cycles) to estimate the spatial and temporal properties of
emerging sunspot groups. Since the relationships we have derived are
only in terms of correlations, our models have a random component.
For the latitudes and areas we draw from random
populations which have the relevant distributions set out in the
previous sections. 

Implicit in such a reconstruction is the assumption that the dynamo has operated
in a similar way from 1700 onwards. The very limited records of observations during the 
earlier part of the 18th century indicate that some of the early cycles might be anomalous
in having stronger activity near the equator than those of the better observed later cycles
\citep[\egc , see][]{Ribes93, Arlt09}. This could indicate that the dynamo was operating in
a not purely dipole mode during this period.

\subsection{A semi-synthetic butterfly diagram covering 1874--2010}
\begin{figure}
\centering
\resizebox{\hsize}{!}{\includegraphics[width=0.7\textwidth,angle=90]{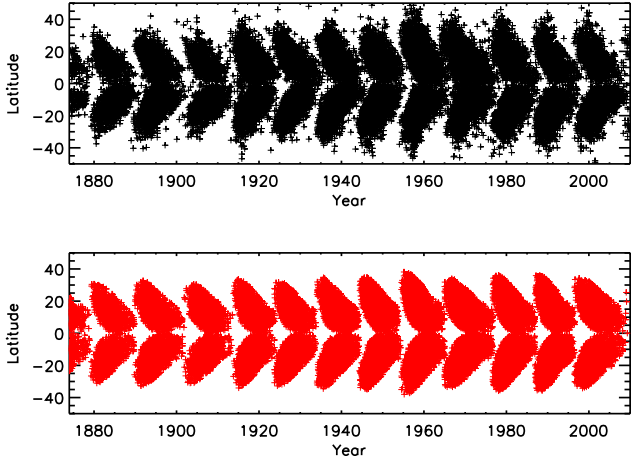}}
\caption{A comparison of observed (upper panel)
and semi-synthetic (lower panel) butterfly diagrams for the years
1874--2010.}\label{fig:butterly_tmin}
\end{figure}

\begin{figure}
\centering
\resizebox{\hsize}{!}{\includegraphics[width=0.5\textwidth,angle=90]{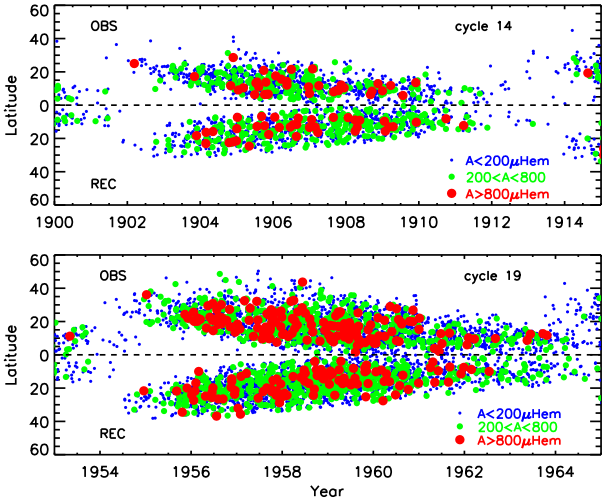}}
\caption{Comparison of butterfly diagrams from observation (above the horizontal dashed lines)
and reconstruction (below the dashed lines) for the weakest cycle 14 (upper panel)
and the strongest cycle 19 (lower panel), both for the northern hemisphere. The
area of the sunspot groups is indicated by the colors and sizes of circles.} \label{fig:butterly_tmin2}
\end{figure}

We first present an example semi-synthetic butterfly
diagram for the period from the start of the RGO records to 2010.
This allows us to directly compare the semi-synthetic and observed
butterfly diagrams in Figure~\ref{fig:butterly_tmin}. As expected, 
the two diagrams have similar appearances. 
A more detailed comparison of the weakest and strongest cycles is 
shown in Figure \ref{fig:butterly_tmin2}. Again the observed and semi-synthetic
butterfly wings look similar. This validates the use of the semi-synthetic
reconstruction for periods when we only have the sunspot numbers.

\subsection{Comparison of reconstructions using $R_G$ and $R_Z$ during 1700--1874}

\begin{figure}
\centering
\resizebox{\hsize}{!}{\rotatebox{90}{\includegraphics{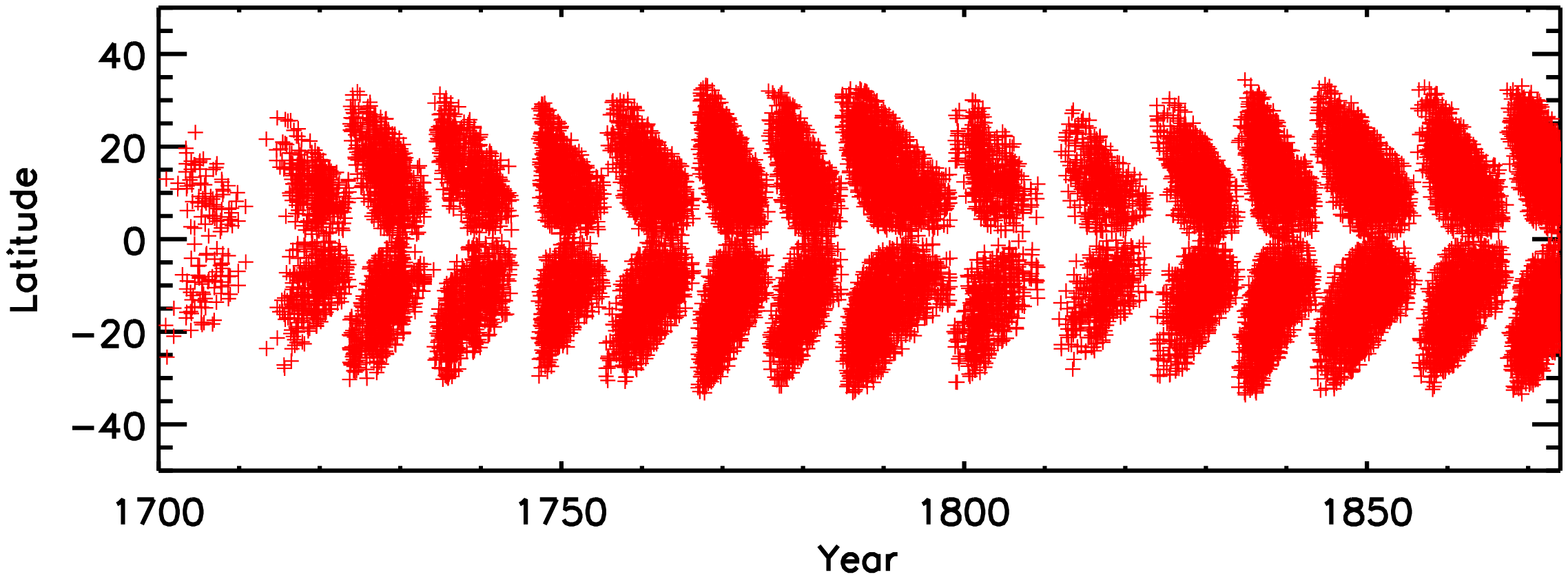}}}
\resizebox{\hsize}{!}{\rotatebox{90}{\includegraphics{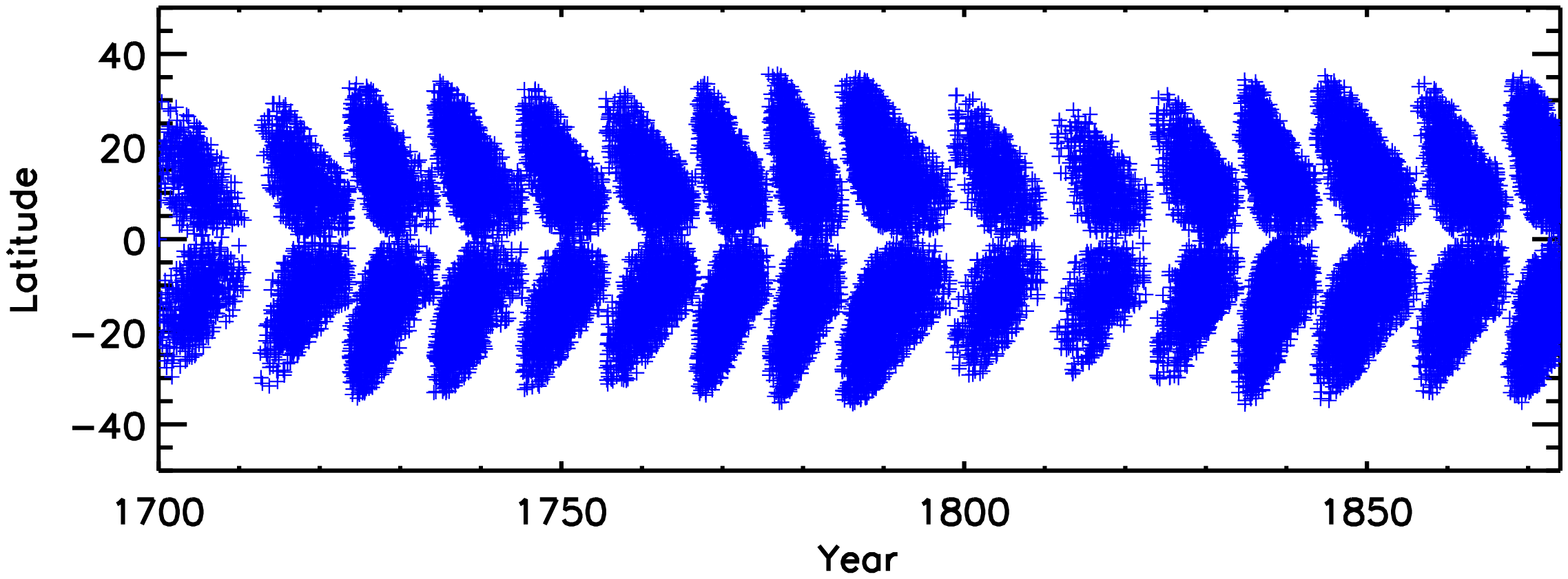}}}
\caption{Semi-synthetic butterfly diagram for the years 1700--1874
using $R_G$ (upper panel) and $R_Z$ (lower panel).}
\label{fig:butterly_both}
\end{figure}

The semi-synthetic model shown in Figures~\ref{fig:butterly_tmin} and \ref{fig:butterly_tmin2}  
was based on the group sunspot number $R_G$. Prior to 1874 $R_Z$ and $R_G$ have substantial differences
which affect the reconstructed butterfly diagrams. Figure \ref{fig:butterly_both}
shows the reconstructed butterfly diagram during 1700--1874 with
$R_G$ and $R_Z$, respectively. It will be very interesting to compare 
both semi-synthetic butterfly diagrams with those being obtained by
\cite{Arlt10}. 
We comment that there is no reason emerging from this study to prefer one 
data set over the other. 

\begin{figure}
\centering
\resizebox{\hsize}{!}{\rotatebox{90}{\includegraphics{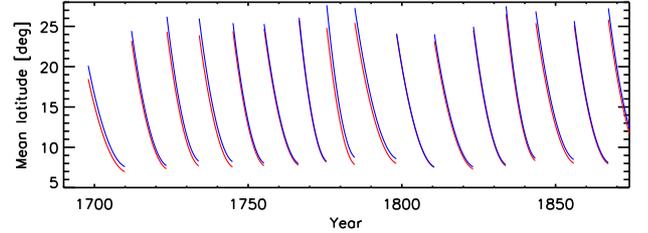}}}
\caption{Mean sunspot latitudes for the years 1700--1874
reconstructed with $R_Z$ (blue curves) and $R_G$ (red curves).}
\label{fig:lati_analy_wolfgroup}
\end{figure}

To give another indication of the differences in the reconstructions based on
$R_G$ and $R_Z$ Figure~\ref{fig:lati_analy_wolfgroup} shows the
reconstructed mean latitudes during 1700--1874. The different cycle strengths 
derived from the two sets of sunspot numbers produce small differences
which differ in strength from cycle to cycle. The extent to which these 
differences affect the results of surface flux transport simulations
and the open flux calculated therefrom will be investigated in Paper II.

\section{Conclusions}

Using the group sunspot number $R_G$ and  RGO, MWO and Kodaikanal data sets, 
we studied the phase dependence and cycle dependence of latitude, area and
tilt angle distribution properties of sunspot group emergence. The main
correlations found are:

1. The mean latitude at which sunspots emerge can be modeled using a second
order polynomial of cycle phase.

2. Strong cycles have a higher mean latitude for sunspot emergence (Figure~\ref{fig:aveLat_str}).

3. The ratio of the latitudinal range over which sunspot groups
emerge and the average latitude of emergence varies as a function of
cycle phase (Figure~\ref{fig:lati_tmin_sca}).

4. The distribution of sunspot areas is similar for all cycles 
(Figure \ref{fig:size_func_cycs}).

5. The size distribution is a power-law for small sunspots and obeys a
log-normal profile for large sunspots (Figure~\ref{fig:size_func_cycs}).

6. During cycle maxima sunspots are, in the mean, larger
(Figure~\ref{fig:area_time_tmin}).

7. Sunspot nests are important, especially during cycle maximum
phases (Figure~6).

We have modeled and used the correlations to construct semi-synthetic butterfly diagrams extending 
back to 1700. 
This reconstruction will be useful in modeling the large-scale solar magnetic 
field over this period.

\bibliographystyle{aa.bst}
\bibliography{16167_bib}

\end{document}